# Extended sample size calculations for evaluation of prediction models using a threshold for classification

Rebecca Whittle[1,2]*, Joie Ensor[1,2], Lucinda Archer[1,2], Gary S. Collins[3], Paula Dhiman[3], Alastair Denniston[2], Joseph Alderman[2,4], Amardeep Legha[1,2], Maarten van Smeden[5], Karel G. Moons[5], Jean-Baptiste Cazier[6], Richard D. Riley[1,2]†, Kym I.E. Snell[1,2]†

**Contact details:**

* Corresponding author

† Joint senior authors

[1] Institute of Applied Health Research, College of Medical and Dental Sciences, University of Birmingham, Birmingham, UK.

[2] National Institute for Health and Care Research (NIHR) Birmingham Biomedical Research Centre, UK.

[3] Centre for Statistics in Medicine, Nuffield Department of Orthopaedics, Rheumatology and Musculoskeletal Sciences, University of Oxford, Oxford, OX3 7LD, United Kingdom

[4] Institute of Inflammation and Ageing, University of Birmingham, Birmingham, UK

[5] Julius Center for Health Sciences and Primary Care, University Medical Centre Utrecht, Utrecht University, Utrecht, The Netherlands.

[6] Francis Crick Institute, London, UK.

**Funding**: This paper presents independent research supported by an EPSRC grant for 'Artificial intelligence innovation to accelerate health research' (number: EP/Y018516/1), and by the National Institute for Health and Care Research (NIHR) Birmingham Biomedical Research Centre at the University Hospitals Birmingham NHS Foundation Trust and the University of Birmingham. GSC is supported by Cancer Research UK (programme grant: C49297/A27294). PD is supported by Cancer Research UK (project grant: PRCPJT-Nov21\100021). RDR, GSC and AD are National Institute for Health and Care Research (NIHR) Senior Investigators. The views expressed in this article are those of the author(s) and not necessarily those of the NHS, the NIHR or the Department of Health and Social Care.

**Competing Interests:** RDR receives royalties for the textbooks 'Prognosis Research in Healthcare' and 'Individual Participant Data Meta-Analysis'.

# Abstract

When evaluating the performance of a model for individualised risk prediction, the sample size needs to be large enough to precisely estimate the performance measures of interest. Current sample size guidance is based on precisely estimating calibration, discrimination, and net benefit, which should be the first stage of calculating the minimum required sample size. However, when a clinically important threshold is used for classification, other performance measures can also be used. We extend the previously published guidance to precisely estimate threshold-based performance measures. We have developed closed-form solutions to estimate the sample size required to target sufficiently precise estimates of accuracy, specificity, sensitivity, PPV, NPV, and F1-score in an external evaluation study of a prediction model with a binary outcome. This approach requires the user to pre-specify the target standard error and the expected value for each performance measure. We describe how the sample size formulae were derived and demonstrate their use in an example. Extension to time-to-event outcomes is also considered. In our examples, the minimum sample size required was lower than that required to precisely estimate the calibration slope, and we expect this would most often be the case. Our formulae, along with corresponding Python code and updated R and Stata commands (*pmvalsampsize*), enable researchers to calculate the minimum sample size needed to precisely estimate threshold-based performance measures in an external evaluation study. These criteria should be used alongside previously published criteria to precisely estimate the calibration, discrimination, and net-benefit.

# 1. Introduction

Prediction models use multiple predictors (features) to provide a prediction of an individual's probability of having (diagnosis) or developing (prognosis) a particular outcome, which can aid health care providers and patients in their decision making. Each year in the medical literature, thousands of prediction models are being reported. But before these prediction models can be considered for use in clinical practice, it is important that the predictive performance of the model is evaluated. This should be done in new data, i.e., in individuals that were not included in the development of the model, that are representative of the target population in whom the model is to be used [1]. As the aim of such evaluation studies is to provide evidence of accuracy of the model's predictions in a particular population, the sample size of an evaluation study needs to be large enough to enable precise estimation of the performance measures of interest [2].

Previous work provides a comprehensive overview of the measures that should be used to assess the performance of a prediction model during an evaluation study [3], and how to calculate the minimum sample size required for evaluation of a prediction model for continuous [4], binary [5,6] and time-to-event outcomes [7]. For binary outcomes, this work focusses on targeting precise estimates of calibration, discrimination, and net benefit measures, which should be the starting point for determining the sample size required to externally evaluate a prediction model.

The underlying goal of many prediction models is to stratify the risk of the patient to help with decision making and treatment strategies. Here we extend these sample size calculations particularly for cases where people are interested in the performance of the model at a particular threshold, i.e. for classification at a clinically important probability threshold. Threshold-based performance measures are not without limitation and need a clear justification to be used in the evaluation of a prediction model, however the recent increase in the use of machine learning methods to develop prediction models has meant there has been a large increase in the use of threshold-based performance measures (although these measures are not restricted to the evaluation of models developed using machine learning methods). Therefore, in this article, we propose closed-form solutions to calculate the minimum sample size required for precisely estimating threshold-based performance measures in an evaluation study of a prediction model that includes an objective to focus on the performance at a threshold. These sample size calculations aim to extend and complement previously published criteria and should be used in addition to sample size calculations for the key performance measures of calibration, discrimination and net benefit discussed by Riley et al. [2,3,5,7]. Recent systematic reviews of prediction models developed using machine learning methods found that accuracy, specificity, sensitivity (recall), Positive Predictive Value [PPV] (precision), Negative Predictive Value [NPV], and the F1-score were common measures of model performance used to evaluate models when using machine learning techniques for model development [8–10] and therefore these are the measure we focus on in this article.

We also investigate whether the previously published sample size calculations would provide sufficient data to precisely estimate these additional measures of performance, or if additional participants would be needed. An example is provided to estimate the width of the confidence interval for each of these additional performance measures if the sample size were equal to that of the minimum required sample size identified using the criteria provided by Riley et al. [2,5,7].

The outline of the paper is as follows. In section 2 an overview of the existing sample size criteria is given. In section 3 the formulae for the threshold-based performance measures of interest are provided, along with formulae to calculate the standard error for each of these measures. An approach to assess the minimum sample size needed to target a particular standard error or confidence interval width for each measure is derived. Section 4 then provides an example of applying these formulae to a prediction model with a binary outcome. In section 5, we extend this for time-to-event outcomes and provide an example of using the formulae in a simulation-based approach to estimate the width of the confidence interval for each of the measures for a given sample size. Code for the examples is available from https://github.com/RebeccaWhittle and can be replicated for the binary example using *pmvalsampsize* in R and Stata [11,12].

## 2. Existing sample size criteria to precisely estimate calibration, discrimination, and net benefit

The previously published approach to calculating the minimum required sample size for externally evaluating a prediction model with a binary outcome [5,6] entails targeting precise estimates of calibration (calibration slope and observed/expected ratio), discrimination (c-statistic), and net benefit. This requires the user to specify the assumed true outcome event proportion and the linear predictor distribution in the external evaluation population, the target standard error of the log of the observed/expected ratio, the target standard error of the calibration slope, the anticipated true c statistic in the external evaluation population, and the probability threshold of interest for clinical decision making, if relevant.

The command *pmvalsampsize* was made available in R and Stata to calculate the minimum required sample size for each criterion based on the required information as stated above [11,12]. The corresponding command in Python will soon be available too.

## 2.1. Criterion 1: sample size to target a precise estimate of the observed/expected ratio (O/E)

To target a precise estimate of the observed/expected (O/E) ratio, we can apply the following equation:

$$N = \frac{1 - \emptyset}{\emptyset \left( SE \left( \ln \left( \frac{O}{E} \right) \right) \right)^2},$$

where $\emptyset$ is the assumed true outcome event proportion in external evaluation population and $SE \left( \ln \left( \frac{O}{E} \right) \right)$ is the target standard error of the $\ln \left( \frac{O}{E} \right)$ statistic, which would ensure a precise confidence interval width on the O/E scale. The choice of $SE \left( \ln \left( \frac{O}{E} \right) \right)$ is context specific because it depends on the overall event probability in the population.

## 2.2. Criterion 2: sample size to target a precise estimate of the calibration slope ($\beta$)

To target a precise estimate of the calibration slope ($\beta$), we can apply the following equation:

$$N = \frac{I_\alpha}{SE(\beta)^2 \left( I_\alpha I_\beta - I_{\alpha\beta}^2 \right)},$$

where $SE(\beta)$ is the target standard error for the calibration slope estimate. The $I_\alpha$, $I_{\alpha\beta}$, and $I_\beta$ are elements of Fisher's information matrix and depend on the distribution of the linear predictor values ($LP_i$, which are the predicted values on the log-odds scale) in the external evaluation population, and correspond to the mean value of $a_i$, mean value of $b_i$, and mean value of $c_i$, respectively, where;

$$a_i = \frac{\exp(\alpha + \beta\ \text{LP}_i)}{(1 + \exp(\alpha + \beta\ \text{LP}_i))^2} \quad b_i = \frac{\text{LP}_i \exp(\alpha + \beta\ \text{LP}_i)}{(1 + \exp(\alpha + \beta\ \text{LP}_i))^2} \quad c_i$$

$$= \frac{(\text{LP}_i)^2 \exp(\alpha + \beta\ \text{LP}_i)}{(1 + \exp(\alpha + \beta\ \text{LP}_i))^2}$$

The R and Stata command *pmvalsampsize* automatically calculates $I_\alpha$, $I_{\alpha\beta}$, and $I_\beta$ based on the specified distribution for $\text{LP}_i$ and the assumed calibration performance [11,12].

### 2.3. Criterion 3: sample size to target a precise estimate of the c statistic

To target a precise estimate of the c statistic, we use the following formula for the standard error of the c statistic, proposed by Newcombe [13], which makes no assumptions about the underlying distribution of the prediction models linear predictor:

$$SE(C) = \sqrt{\frac{C(1-C)\left(1 + \left(\frac{N}{2} - 1\right)\left(\frac{1-C}{2-C}\right) + \frac{\left(\frac{N}{2} - 1\right)C}{1+C}\right)}{N^2\emptyset(1-\emptyset)}}.$$

Here, $C$ is the anticipated true c statistic for the external evaluation population, $\emptyset$ is the assumed true outcome event proportion in the external evaluation dataset, and $SE(C)$ is the target standard error of C. Given these values, the R and Stata command *pmvalsampsize* uses an iterative process to identify the minimum required sample size, $N$ [11,12].

### 2.4. Criterion 4: sample size to target a precise estimate of the standardised net benefit (sNB) at a clinically important probability threshold ($p$)

To target a precise estimate of the standardised net benefit (sNB) at a clinically important probability threshold ($p$), we apply the following equation derived by Marsh et al. [14]:

$$N = \frac{1}{SE(sNB)^2}\left(\frac{sensitivity(1 - sensitivity)}{\emptyset} + \frac{w^2 specificity(1 - specificity)}{1 - \emptyset}\right.$$
$$\left. + \frac{w^2(1 - specificty)^2}{\emptyset(1 - \emptyset)}\right),$$

where $sNB$ is defined as $NB/\emptyset$, based on sensitivity and specificity values that correspond to the chosen distribution of predicted values and assuming the model is well calibrated. Also,

$w = \frac{1-\emptyset}{\emptyset} \frac{p}{1-p}$, and the models sensitivity and specificity are the anticipated values at threshold $p$.

## 3. Sample size calculations for performance measures commonly used when evaluating classification-based prediction models

Consider that one wishes to evaluate the performance of an existing prediction model that has a binary outcome (i.e., $Y = 0$ or $1$) and that the model can be used to calculate the probability of the outcome event or a score for an individual participant. If the model has been developed via a machine learning approach, it may not have a related equation as such, but can still be validated if it outputs estimated probabilities for each individual in the evaluation dataset. For external evaluation, we require a representative dataset containing $N$ participants from the target population of interest, who were not included in the data used to develop the model. We need to make sure $N$ is large enough to estimate the model's predictive performance precisely, and we now introduce key terminology to do this.

Firstly, let us assume a certain probability threshold, $p$, used to define the predicted presence or absence of the outcome. Those with a predicted probability above this threshold will be classified as positive, or predicted to have the outcome, and those with a predicted probability lower than this threshold will be classified as negative, or predicted to not have the outcome. The number of true positives (TP) can be calculated as the number of participants who had a predicted probability $> p$ and had the outcome; false positives (FP) can be calculated as the number of participants that had a predicted probability $> p$ but did not have the outcome; true negatives (TN) will be the number with a predicted probability $< p$ and didn't have the outcome; and false negatives (FN) can be calculated as the number of participants who had a predicted probability $< p$ but did have the outcome. Then, the total sample size is $N = TP + FP + TN + FN$, as can be seen in the confusion matrix below:

|  | Outcome | No outcome |  |
|---|---|---|---|
| **Prob>$p$** | TP | FP | *TP+FP* |
| **Prob<$p$** | FN | TN | *FN+TN* |
|  | *TP+FN* | *FP+TN* | **N** |

The outcome proportion, i.e., the number of participants in the sample with the outcome, can be defined as:

$$\emptyset = \frac{TP + FN}{N}. \tag{1}$$

These values (TP, FP, TN, FN, N, and $\emptyset$) are used in the calculation of our additional performance measures and their associated standard errors. Given this, we now derive a formula to calculate the minimum required sample size ($N$) to achieve a specified width of the confidence interval for each measure: accuracy, specificity, sensitivity, PPV, NPV and F1-Score. Each of the formulae for the performance measures, their standard errors, and the sample size required to achieve adequate precision, are provided in Table 1.

### 3.1. Classification accuracy

Classification accuracy is a commonly used measure of model performance for models developed using machine learning techniques (however, it also applies to models developed using statistical methods). It is defined as the proportion of correct classifications the model has made, based on a particular threshold $p$. The definition of a 'good' model based on its accuracy is context specific and depends on what the model is predicting.

The accuracy can be calculated as:

$$Accuracy = \frac{Number\ of\ correct\ predictions}{Total\ number\ of\ predictions} = \frac{TP + TN}{N}. \tag{2}$$

Using the formula for the standard error of a proportion, the standard error of the estimate of the accuracy is defined approximately as,

$$SE_{acc} = \sqrt{\frac{accuracy \times (1 - accuracy)}{N}}, \tag{3}$$

which can be rearranged to obtain the minimum sample size required to target a pre-specified standard error of the accuracy using an assumed true value of the accuracy:

$$N = \frac{accuracy \times (1 - accuracy)}{SE_{acc}^2}. \tag{4}$$

Alternatively, $N$ can be calculated using a pre-specified target confidence interval width ($CIW_{acc}$) rather than standard error:

$$N = \frac{accuracy \times (1 - accuracy)}{\left(\frac{CIW_{acc}}{2 \times 1.96}\right)^2}, \qquad (5)$$

assuming the error is normally distributed and so using a confidence interval of the form:

$$accuracy \pm 1.96 \times SE_{acc}^2.$$

## 3.2. Specificity

Specificity is a measure of the models' ability to correctly predict true negatives, i.e., the proportion of individuals without the outcome that the model predicts to not have the outcome. The specificity can be calculated as:

$$Specificity = \frac{TN}{TN + FP}. \qquad (6)$$

The standard error for the specificity is then defined approximately as,

$$SE_{spec} = \sqrt{\frac{specificity \times (1 - specificity)}{TN + FP}}. \qquad (7)$$

Recognising $TN + FP = N(1 - \emptyset)$, the formula can be rearranged to calculate the minimum required sample size for a pre-specified target standard error, based on the expected outcome proportion and an assumed true value of the specificity:

$$N = \frac{specificity \times (1 - specificity)}{SE_{spec}^2 \times (1 - \emptyset)}. \qquad (8)$$

This can be re-written in terms of the target CI width ($CIW_{spec}$):

$$N = \frac{specificity \times (1 - specificity)}{\left(\frac{CIW_{spec}}{2 \times 1.96}\right)^2 \times (1 - \emptyset)}. \qquad (9)$$

## 3.3. Sensitivity

Sensitivity, or recall as it is commonly called in the machine learning literature, measures how often the model identifies individuals with the outcome from the total individuals with the outcome in the sample, i.e., is a measure of the models' ability to predict true positives. If we had a sensitivity of say 0.6, this would be interpreted as the model correctly identifying 60% of all participants with the outcome as positive. As the threshold for classifying a prediction as positive ($p$) increases, the number of false negatives will increase, and so the sensitivity decreases.

The sensitivity can be calculated as:

$$Sensitivity = \frac{TP}{TP + FN}. \tag{10}$$

The standard error of the sensitivity is therefore defined as approximately:

$$SE_{sens} = \sqrt{\frac{sensitivity \times (1 - sensitivity)}{TP + FN}}, \tag{11}$$

which can be rearranged (recognising $TP + FN = N\emptyset$) to obtain the following formula for the sample size, $N$, conditional on a pre-specified target standard error, an expected outcome proportion, and an assumed true value of the sensitivity:

$$N = \frac{sensitivity \times (1 - sensitivity)}{\emptyset \times SE_{sens}^2}, \tag{12}$$

and can also be written in terms of the target CI width of the sensitivity:

$$N = \frac{sensitivity \times (1 - sensitivity)}{\emptyset \times \left(\frac{CIW_{sens}}{2 \times 1.96}\right)^2}. \tag{13}$$

## 3.4. Positive Predictive Value (PPV)

Positive Predictive Value (PPV), commonly known as precision in the machine learning literature, is a measure of quality of the predictions from the model. It measures the proportion of individuals that are predicted to have the outcome that do have the outcome. For example, if the PPV was 0.7, then when the model predicts someone to have the outcome, it is correct 70% of the time. As the threshold for classifying as positive ($p$) increases, the number of false positives decreases, and so the PPV will increase.

The precision can be calculated as:

$$PPV = \frac{TP}{TP + FP}. \tag{14}$$

The standard error of the precision is defined as approximately,

$$SE_{PPV} = \sqrt{\frac{PPV \times (1 - PPV)}{TP + FP}}. \tag{15}$$

Recognising that $TP = PPV \times N\emptyset$ and $FP = TP(1 - PPV)/PPV$, and therefore $TP + FP = (sensitivity \times N\emptyset)/PPV$, this can be written as:

$$SE_{PPV} = \sqrt{\frac{PPV^2 \times (1 - PPV)}{N \times \emptyset \times sensitivity}},$$

which can be rearranged to get the following formula for $N$ conditional on a pre-specified target standard error, expected outcome proportion, and assumed true values of the PPV and sensitivity,

$$N = \frac{PPV^2 \times (1 - PPV)}{SE_{PPV}^2 \times \emptyset \times sensitivity}. \tag{16}$$

This can be written in terms of the target CI width of the PPV:

$$N = \frac{PPV^2 \times (1 - PPV)}{\left(\frac{CIW_{PPV}}{2 \times 1.96}\right)^2 \times \emptyset \times sensitivity}. \tag{17}$$

### 3.5. Negative Predictive Value (NPV)

The negative predictive value (NPV) measures the proportion of individuals which the model classifies as not having the outcome, who do not have the outcome.

The NPV can be calculated as:

$$NPV = \frac{TN}{TN + FN}, \tag{18}$$

with its standard error approximately defined as:

$$SE_{NPV} = \sqrt{\frac{NPV \times (1 - NPV)}{TN + FN}}, \tag{19}$$

where $TN = N \times specificity \times (1 - \emptyset)$ and $FN = N\emptyset(1 - sensitivity)$, and hence the sample size, $N$, can be written in terms of assumed true values of NPV, specificity, and sensitivity, the expected outcome proportion, and a pre-specified target standard error for the NPV:

$$N = \frac{NPV \times (1 - NPV)}{SE_{NPV}^2 \times (specificity \times (1 - \emptyset) + \emptyset \times (1 - sensitivity))}, \tag{20}$$

which can be written in terms of a target CI width of the NPV:

$$N = \frac{NPV \times (1 - NPV)}{\left(\frac{CIW_{NPV}}{2 \times 1.96}\right)^2 \times (specificity \times (1 - \emptyset) + \emptyset \times (1 - sensitivity))}. \tag{21}$$

### 3.6. F1-score

As the threshold ($p$) increases, the number of false negatives increases and the number of false positives decreases, meaning that the sensitivity will decrease as the PPV increases. Hence, PPV and sensitivity compete with each other. Due to this, there have been scores developed that encompass both the PPV and sensitivity, and a popular score that is used for evaluating machine learning models is the F1-score. In the machine learning literature, sensitivity is usually referred to as recall, and PPV as precision, therefore when discussing the F1-score in this article, sensitivity and PPV will be referred to as recall and precision.

The F1-score is a measure of the model's accuracy which combines the precision and recall using the harmonic mean. Ranging from 0 to 1, it measures how many times a model makes a correct prediction across the entire dataset, with 1 indicating that the model accurately predicted every individual's outcome. The harmonic mean incorporates the reciprocal of each value, and so is more highly weighed towards the smaller number in the pair, unlike the arithmetic mean. Thus, the F1-score will only be high when both precision and recall are high, ensuring a balance of both aspects of model performance. The F1-score is often used instead of accuracy when there is severe class imbalance in the dataset.

As with other measures of performance, a 'good' score depends on what the model is predicting, e.g., sometimes recall is more important than precision, even at the cost of precision, and vice versa, but the F1-score aims to give a balance of the precision and recall.

The F1-score can be calculated as the harmonic mean of the precision and recall:

$$F1 = \frac{2}{\frac{1}{precision} + \frac{1}{recall}} = 2 \times \frac{precision \times recall}{precision + recall}$$

$$= \frac{2}{\frac{TP + FP}{TP} + \frac{TP + FN}{TP}}. \tag{22}$$

Using the formula provided by Takahashi et al [15] for the variance of a macro-averaged F1-score (the average of F1-scores in each class of a multi-class classification model), and amending for the binary scenario, the standard error of the F1-score can be defined as:

$$SE_{F1} = \sqrt{4 \times \frac{R^4 SE_P^2 + 2P^2 R^2 Cov(P, R) + P^4 SE_R^2}{(P + R)^4}}, \tag{23}$$

where $P$ is the precision and $R$ is the recall, and the covariance of the precision and recall is:

$$cov(P, R) = \frac{FP \times TP \times FN}{(TP + FP)^2 (TP + FN)^2} + \frac{FP \times TP \times TN}{(TP + FP)^2 (TN + FP)^2}$$

$$= \frac{\frac{P(1-P)(1-R)}{\emptyset} + \frac{P(1-P) \times Specificity}{1 - \emptyset}}{N}. \tag{24}$$

Eq. (23) and (24) can then be rearranged to obtain a solution for the sample size, $N$, conditional on pre-specified target standard errors for precision and recall, expected outcome proportion, and assumed true values of the precision, recall, and specificity:

$$N = \frac{2P^2R^2\left(\frac{P(1-P)(1-R)}{\emptyset} + \frac{P(1-P) \times Specificity}{1-\emptyset}\right)}{\left(\frac{SE_{F1}^2(P+R)^4}{4}\right) - R^4SE_P^2 - P^4SE_R^2}, \quad (25)$$

which can also be expressed in terms of the target CI widths:

$$N = \frac{2P^2R^2\left(\frac{P(1-P)(1-R)}{\emptyset} + \frac{P(1-P) \times Specificity}{1-\emptyset}\right)}{\left(\frac{\left(\frac{CIW_{F1}}{2 \times 1.96}\right)^2 (P+R)^4}{4}\right) - R^4\left(\frac{CIW_P}{2 \times 1.96}\right)^2 - P^4\left(\frac{CIW_R}{2 \times 1.96}\right)^2}. \quad (26)$$

**3.7. Implementing the sample size calculations**

The approach requires the user to pre-specify the target standard error (or confidence interval width) that is their maximum desired. This is subjective, but generally we suggest a confidence interval width of 0.1 or less, because this represents good precision on the 0 to 1 (proportion) scale of all these classification-based measures. The calculations also require the anticipated value of the outcome proportion and the performance measures to be specified. We suggest these are based on the performance estimates shown in the original publication of the developed model, especially if we would expect the validation sample to be from the same population as that used for the development sample, and the model to have good calibration.

## 4. Applied example: external validation of a prognostic model for predicting deterioration in hospitalised adults with COVID-19

Here we extend the applied example provided by Riley et al. [2] calculating the minimum required sample size to externally validate the 'ISARIC 4C Deterioration model' which was developed by Gupta et al. in 2021 [16]. It is a multivariable logistic regression model for predicting in-hospital clinical deterioration (defined as any requirement of ventilatory support or critical care, or death) among hospitalised adults with highly suspected or confirmed COVID-19. The model was developed using data from 260 hospitals including 66,705 participants across England, Scotland, and Wales, and evaluated in a separate dataset of 8,239 participants from London. Initial model evaluation was judged satisfactory and Riley et al.[2] provided sample size calculations to further externally evaluate the model to check predictions were still reliable following the introduction of COVID-19 vaccines and other interventions.

Assuming the model would be well calibrated, with an anticipated c statistic of 0.77 (based on the previous evaluation study), and assuming the distribution of the model's predicted event probabilities in the external evaluation population would be similar to that in the histogram presented by Gupta et al. in their supplementary material [16], Riley et al.[2] approximated this by using a Beta(1.33, 1.75) distribution (Figure 1), which yielded a similar shape and had the same overall outcome event proportion of 0.43 as in the previous evaluation study. Riley et al.[2] calculated the minimum sample size required to precisely estimate the calibration, discrimination and net benefit. Here we further extend this to consider what the minimum required sample size would be if we were also interested in evaluating the additional performance measures described in this article: accuracy, specificity, sensitivity, PPV, NPV and F1-score.

### 4.1. Calculating the minimum required sample size to target a particular confidence interval width

First we simulated a large dataset of 1,000,000 observations with outcome probabilities approximated by a Beta(1.33, 1.75) distribution. From this simulated data we can calculate the 'true' accuracy, specificity, sensitivity, PPV, NPV, and F1-score that would be anticipated in the new external evaluation study. Table 2 provides the 'true' value of the performance measure along with minimum required sample sizes calculated using the formulae in Table 1, at threshold probabilities of 0.1 and 0.3 (as evaluated in the original article), and also at different target CI widths (0.08, 0.1 and 0.12).

When targeting a confidence interval width of 0.1 for each of the six performance measures, (i.e., an assumed standard error of 0.0255), the minimum required sample size is calculated as 385, 338, 42, 423, 933 and 379 for accuracy, specificity, sensitivity, PPV, NPV, and F1-score, respectively (Table 2). Hence, based on these measures alone, the minimum required sample size would be 933 (402 events) which is driven in this case by the sample size needed to precisely estimate the NPV.

### 4.2. Comparison to the sample size needed for calibration, discrimination, and utility

We now compare these calculated sample sizes to those previously derived for other measures (Table 3). For their planned external evaluation study, Riley et al. targeted a confidence interval width of 0.22 for the O/E statistic, 0.3 for the calibration slope, 0.1 for the c statistic, and 0.2 for the standardised net benefit, which gave a minimum required sample size (number of events) of 423 (182) for O/E, 949 (408) for calibration slope, 347 (149) for the c statistic, and 38 (16) for the standardised net benefit at a threshold of 0.1 and 407 (175) for the standardised net benefit at a threshold of 0.3. Therefore, at least 949 participants (408 events) would be required, which is driven by the sample size needed to precisely estimate the calibration slope. In this example, this is larger than the sample size needed for the measures introduced in this article.

### 4.3. Calculating the expected confidence interval width based on a given sample size

As an alternative to targeting a particular confidence interval width for the performance measures, it is possible to calculate the expected width of the confidence intervals based on a given sample size. Table 4 provides the performance values with their corresponding 95% CIs based on a sample size of 949, as calculated for the applied example in Riley et al. [2]. Each of the additional performance measures of interest will be precisely estimated at the minimum required sample size of 949, i.e., the confidence interval widths are all narrower than the 0.1 maximum width specified.

### 4.4. Approximate binomial confidence intervals

The sample size calculations given for accuracy, specificity, sensitivity, PPV and NPV use formula which approximates the distribution of the error for a proportion using a normal distribution, known as the Wald interval. However, this is unreliable when the sample size is small, or when the value of the proportion is close to the extremes (i.e. close to 0 or 1). A better approach may be to use the Agresti-Coull interval [17], which adjusts the simple normal approximation by adding two successes and 2 failures, to give a point estimate defined as:

$$prop_{obs} = \frac{x}{n} \Rightarrow prop_{adj} = \frac{x+2}{n+4},$$

where $x$ is the observed numerator of the proportion and $n$ is the observed denominator of the proportion. The Agresti-Coull confidence interval for the proportion of interest is then calculated as

$$prop_{adj} \pm 1.96 \sqrt{\frac{prop_{adj}(1 - prop_{adj})}{n}}.$$

For example, the confidence interval for sensitivity would be calculated as:

$$\frac{TP + 2}{TP + FN + 4} \pm 1.96 \sqrt{\frac{\left(\frac{TP + 2}{TP + FN + 4}\right) \times \left(1 - \frac{TP + 2}{TP + FN + 4}\right)}{TP + FN}}.$$

Unlike when using the Wald interval, it is not possible to get closed-form solutions for the sample size required when targeting a pre-specified standard error using the Agresti-Coull interval. However, the sample sizes can be obtained by employing an iterative approach. We repeated the example above but using the Agresti-Coull interval and iteratively estimating the minimum required sample size to target a confidence interval width of 0.1. The estimated minimum required sample sizes were 384, 339, 42, 420 and 935 for accuracy, specificity, sensitivity, PPV, and NPV, respectively, for a probability threshold of 0.1. These are negligibly different to those which were calculated using the Wald interval and we expect the choice of interval to make little difference to the estimate of the sample size required in reality. Furthermore, when the sample size calculated is small for a particular measure, it is likely that the sample size required to precisely estimate the calibration slope will be much larger, and so will drive the overall minimum required sample size.

## 5. Extension: Time-to-event outcomes

### 5.1. Situations with no censoring before the time horizon for prediction

When considering a time-to-event scenario, if there is no censoring by the time point of interest, then the outcome event status (i.e., presence or absence of a certain health outcome) is truly known for all individuals in the validation dataset. For a model developed using a proportional hazards regression, an individual's probability of having the outcome event by time $t$ calculated as

$$\hat{F}_i(t) = 1 - \hat{S}_i(t)$$
$$= 1 - \hat{S}_0(t)^{\exp(\text{LP}_i)},$$

where $\hat{S}_i(t)$ is the survival probability by time $t$, and $\hat{S}_0(t)$ is the baseline survival probability at time $t$ and the models linear predictor (LP$_i$) is $\hat{\beta}_i X_{1i} + \hat{\beta}_2 X_{2i} + \cdots + \hat{\beta}_k X_{ki}$. If either $\hat{S}_0(t)$ or $\hat{F}_0(t)$ has been reported in the original model development publication, then absolute risk predictions can be calculated at the time point of interest and the formulae in Table 1 can be used to calculate the performance measures at this time point of interest, and similarly the sample size required to target a pre-specified confidence interval width can be calculated using the formulae in Table 1.

### 5.2. Situations with censoring before the time horizon for prediction

Where there are censored observations before the time point of interest for prediction, then the true outcome event status is unknown for the censored individuals. The use of pseudo-observations (or pseudo-values) has been proposed to address this, which give pseudo-observed event probabilities for each individual accounting for non-informative right censoring, derived using the jackknife estimator [18,19].

The pseudo-observation for individual $i$ is calculated as

$$\tilde{F}_i(t) = N F_{KM}(t) - \left[(N-1) F_{KM(-i)}(t)\right],$$

where $F_{KM}(t) = 1 - S_{KM}(t)$ is the Kaplan-Meier (or another nonparametric) estimate of the cumulative incidence at time $t$ using all $N$ individuals in the validation dataset, and $F_{KM(-i)}(t)$ is the cumulative incidence estimate recalculated on the $N-1$ individuals after removing individual $i$. This allows the confusion matrix to be defined at a given probability threshold to enable the performance measures to be calculated and a simulation-based approach can be used to estimate the anticipated width of the confidence intervals for each of the measures based on a given sample size [7].

### 5.3. Applied example: sample size to externally validate a model for recurrent venous thromboembolism (VTE) following cessation of therapy for a first VTE

Here we extend an applied example provided by Riley et al [7] to identify the required sample size to evaluate a prognostic time-to-event model developed by Ensor et al [20], for 3-year risk of recurrent venous thromboembolism (VTE) following cessation of therapy for a first VTE. It was assumed that the evaluation population would be similar to the development population in terms of survival time and censoring distributions, and using a simulation process to target a confidence interval width of <0.2 for the calibration slope estimate identified a minimum required sample size of 14250 (with approximately 1430 events observed by 3 years), or 3600 individuals (approximately 365 events) to target a confidence interval width of 0.4.

Further details on this simulation process are given in Riley et al. [7] with detailed instructions for each step, along with code for the example used.

We repeat this simulation study, but additionally include the performance measures outlined in Section 2, so to estimate the expected confidence intervals for the estimates of accuracy, specificity, sensitivity, PPV, NPV and the F1-score at various sample sizes.

Hence, we can see that each of the additional performance measures of interest will be precisely estimated at the minimum required sample size of 3600, i.e., the confidence interval widths are all <0.06, and so narrower than the 0.1 maximum width specified.

## 6. Discussion

We have built on previous work to provide additional criteria to be considered when calculating the minimum required sample size for an evaluation of a prediction model when planning to use performance measures typically used in the evaluation of models developed using a threshold for classification (accuracy, specificity, sensitivity, PPV, NPV, F1-score). We have derived closed-form solutions for calculating the minimum sample size required for evaluation of a prediction model with a binary outcome, and illustrated a simulation-based approach to calculating the minimum required sample size for a model using a time-to-event outcome. Prediction models developed using a threshold for classification are rapidly becoming more prevalent, partly due to the recent increase in the use of machine learning methods to develop prediction models, and so in turn, the additional performance measures discussed in this article are also becoming more prevalent.

The additional criteria provided in this article should be used alongside previously published criteria [2,5,7] where there is a known threshold of clinical importance, and it is therefore important to assess the performance at that threshold. Thus, we hope that if researchers are evaluating a model where there is a known threshold of clinical importance, these additional criteria will allow researchers to plan the evaluation study such that the sample size is adequate not only to precisely estimate measures of calibration, discrimination and net benefit, but also any classification-based performance measures deemed suitable for evaluation.

We have provided examples for both a binary and a time-to-event outcome, with Python code available for the binary example and Stata code available for the time-to-event outcome at https://github.com/RebeccaWhittle. The R and Stata command *pmvalsampsize* [11,12] has also been updated to enable the user to calculate the minimum required sample size based on a given confidence interval width for these threshold-based performance measures, or to calculate the expected confidence intervals for a given sample size, i.e., for the minimum required sample size calculated using the Riley et al. criteria [2,5,7].

Confidence intervals for the measures considered in this article are rarely provided in model development or evaluation studies, however, these formulae can still be used in advance of model evaluation to estimate the required sample size to achieve sufficiently precise estimates of the measures of interest. The closed-form solutions provided along with code to implement may also potentially encourage researchers to more often include confidence intervals for these measures in development and evaluation studies.

Defining the target width of the confidence interval for each performance measure is subjective, with appropriate targets potentially being context specific, but will influence the sample size needed and therefore requires careful consideration. The example in this article shows that even a small change of 0.02 to the confidence interval width can have a large impact on the minimum sample size required, which highlights the importance of specifying a sensible target confidence interval width prior to calculating the sample size.

To implement this approach, we are also required to make some assumptions about the distribution of the evaluation data, which can be informed by the development study or by previous evaluation studies. If the data generating assumptions are correct, then the confidence interval width chosen (or estimated at a given sample size) will be achieved on average when the dataset of the identified sample size is generated. However, the confidence interval may be wider or narrower simply by chance, even when the distribution assumptions made were correct. It may also be difficult to define prediction distribution from a model developed using machine learning methods as they are not well described by conventional statistical methods and key information is often not reported. Further guidance is provided by Riley et al. [2].

Most of the formulae presented here are based on the large sample solution for the standard error of a proportion. As the sample sizes required will tend to be large, this is appropriate. However, we have provided Python code at https://github.com/RebeccaWhittle and an option has been added to the *pmvalsampsize* command, in both R and Stata [11,12], to calculate sample sizes iteratively using the Agresti-Coull interval to approximate a binomial confidence interval.

In the simulation procedure for time-to-event outcomes, we have assumed that the censoring is non-informative as a pragmatic decision to ease the computation process. However, in practice censoring may be informative, which should be considered once the evaluation data has been collected. As the pseudo-observations assume non-informative censoring, we suggest that to help mitigate this they can be derived separately within groups defined by tenths or twentieths of predicted risk. Alternatively, Austin et al. suggests to use flexible adaptive hazard regression or a Cox model using restrictive cubic splines instead [21].

There are several limitations of using thresholds for classification that should be noted, and although here we provide solutions to calculate the minimum sample size required to ensure precise estimation of these threshold-based performance measures, we do not endorse using these measures unless there is strong evidence to support their use, i.e. there is clinical/patient input, economic input, or a biological rationale. They should also always be used as an addition to the sample size criteria for discrimination and calibration, rather than instead of.

Defining an appropriate risk threshold for classification is a particular challenge and a plausible risk threshold depends on the clinical context [22]. They are often defined in an ad hoc way, lacking clinical or theoretical foundation, or using a purely statistical criterion which may be inappropriate in practice, and are also subject to sampling variability so a different threshold would be obtained in another dataset from the same population [22]. This clearly will then impact the estimation of the threshold-based performance measures. Threshold based performance measures can also change if the evaluation population is different from the population the model was developed in, which makes it difficult to prespecify what the anticipated values of the measures are, especially if the outcome prevalence is different. These measures will also be likely to vary across subgroups [23]. Moons and Harrell also advise against placing an emphasis on the use of sensitivity and specificity in diagnostic accuracy studies as they are of limited value to clinical practice [24].

If thresholds are being considered, it is recommended to consider a range of acceptable risk thresholds [22], and we would also advise this should also be the case when determining the sample size required to evaluate a model using threshold-based performance measures.

In conclusion, we have provided formulae to enable the standard error, and hence confidence intervals, of the accuracy, specificity, sensitivity, PPV, NPV and the F1-score to be estimated. We have also provided formulae to calculate the minimum sample size required to precisely evaluate these measures in a prediction model for a binary outcome. Further, we have illustrated how these formulae could be used for assessing the minimum sample size needed to precisely evaluate a time-to-event outcome using a simulation-based approach.

In each of our examples, the minimum sample size required was lower than that required to precisely estimate the calibration slope, and we expect this would most often be the case. Therefore, the precise estimation of the calibration slope should remain the focus of any sample size calculation for evaluation of a prediction model, regardless of whether a threshold is being used for classification or of the methods used to develop the model, and the formulae provided in this article should be seen as complementary to the previously published criteria [2,5,7]. We hope that the additional criteria provided, and the worked examples presented in this article, will empower those conducting evaluation studies to make more meaningful judgements regarding sample sizes when considering threshold-based performance measures in addition to calibration, discrimination and net-benefit, improving the quality of model assessments provided in these studies moving forwards.

*Table 1: Formulae to calculate each performance measure, their standard error, and N*

| | | Standard Error | N |
|---|---|---|---|
| **Classification accuracy** | $\dfrac{TP + TN}{N}$ | $\sqrt{\dfrac{accuracy \times (1 - accuracy)}{N}}$ | $\dfrac{accuracy \times (1 - accuracy)}{SE_{acc}^2}$ |
| **Specificity** | $\dfrac{TN}{TN + FP}$ | $\sqrt{\dfrac{specificity \times (1 - specificity)}{TN + FP}}$ | $\dfrac{specificity \times (1 - specificity)}{SE_{spec}^2 \times (1 - \emptyset)}$ |
| **Sensitivity (recall)** | $\dfrac{TP}{TP + FN}$ | $\sqrt{\dfrac{sensitivity \times (1 - sensitivity)}{TP + FN}}$ | $\dfrac{sensitivity \times (1 - sensitivity)}{SE_{sens}^2 \times \emptyset}$ |
| **Positive Predictive Value [PPV] (precision)** | $\dfrac{TP}{TP + FP}$ | $\sqrt{\dfrac{PPV \times (1 - PPV)}{TP + FP}}$ | $\dfrac{PPV^2 \times (1 - PPV)}{SE_{PPV}^2 \times \emptyset \times sensitivity}$ |
| **Negative Predictive Value [NPV]** | $\dfrac{TN}{TN + FN}$ | $\sqrt{\dfrac{NPV \times (1 - NPV)}{TN + FN}}$ | $\dfrac{NPV \times (1 - NPV)}{SE_{NPV}^2 \times (specificity \times (1 - \emptyset) + \emptyset \times (1 - sensitivity))}$ |
| **F1-score** | $\dfrac{2}{\dfrac{1}{precision} + \dfrac{1}{recall}}$ | $\sqrt{4 \times \dfrac{R^4 SE_P^2 + 2P^2 R^2 Cov(P,R) + P^4 SE_R^2}{(P + R)^4}}$ where $cov(P,R) = \dfrac{\dfrac{P(1-P)(1-R)}{\emptyset} + \dfrac{P(1-P) \times Specificity}{1 - \emptyset}}{N}$ | $\dfrac{2P^2 R^2 \left(\dfrac{P(1-P)(1-R)}{\emptyset} + \dfrac{P(1-P) \times Specificity}{1 - \emptyset}\right)}{\left(\dfrac{SE_{F1}^2 (P + R)^4}{4}\right) - R^4 SE_P^2 - P^4 SE_R^2}$ |

*Table 2: Sample size calculations for the external validation of ISARIC 4C Deterioration model based on targeting precise estimates of accuracy, specificity, sensitivity, PPV, NPV and F1-score. Sample sizes are reported as the minimum total participants (number of events) required.*

|  | Threshold probability | Performance measure value | CI width | | |
|---|---|---|---|---|---|
|  |  |  | 0.08 | 0.1 | 0.12 |
|  |  |  | N (events) | | |
| **Accuracy** | 0.1 | 0.51 | 601 (259) | 385 (166) | 267 (115) |
|  | 0.3 | 0.663 | 537 (231) | 344 (148) | 239 (103) |
| **Specificity** | 0.1 | 0.147 | 529 (228) | 338 (146) | 235 (102) |
|  | 0.3 | 0.508 | 1054 (454) | 675 (291) | 469 (202) |
| **Sensitivity** | 0.1 | 0.988 | 65 (28) | 42 (19) | 29 (13) |
|  | 0.3 | 0.867 | 644 (277) | 413 (178) | 287 (124) |
| **PPV** | 0.1 | 0.468 | 660 (284) | 423 (182) | 294 (127) |
|  | 0.3 | 0.573 | 904 (389) | 579 (249) | 402 (173) |
| **NPV** | 0.1 | 0.943 | 1457 (627) | 933 (402) | 648 (279) |
|  | 0.3 | 0.834 | 959 (413) | 614 (265) | 426 (184) |
| **F1-score** | 0.1 | 0.636 | 592 (255) | 379 (163) | 263 (114) |
|  | 0.3 | 0.690 | 868 (374) | 555 (239) | 386 (166) |

*Table 3: Sample size calculations for the external validation of ISARIC 4C Deterioration model based on Riley criteria*

|  | Threshold probability | Performance measure value | CI Width | N (events) |
|---|---|---|---|---|
| O/E | - | 1 | 0.22 | 423 (182) |
| Calibration slope | - | 1 | 0.3 | 949 (408) |
| C statistic | - | 0.77 | 0.1 | 347 (149) |
| Net benefit | 0.1 | 0.371 | 0.2 | 38 (16) |
|  | 0.3 | 0.216 | 0.2 | 407 (175) |

*Table 4: Estimated performance (95% CI) for accuracy, specificity, sensitivity, PPV, NPV and F1-score at threshold probabilities 0.1 and 0.3 when evaluating the ISARIC 4C Deterioration model, when N=949 (calculated by the Riley criteria)*

|  | **Threshold probability** | |
|---|---|---|
|  | **0.1** | **0.3** |
| **Accuracy** | 0.510 (0.478, 0.542) | 0.663 (0.636, 0.693) |
| **Specificity** | 0.147 (0.117, 0.177) | 0.508 (0.466, 0.550) |
| **Sensitivity** | 0.988 (0.977, 0.999) | 0.867 (0.834, 0.900) |
| **PPV** | 0.468 (0.435, 0.501) | 0.573 (0.534, 0.612) |
| **NPV** | 0.943 (0.894, 0.992) | 0.834 (0.794, 0.874) |
| **F1-score** | 0.636 (0.603, 0.668) | 0.690 (0.652, 0.728) |

*Table 5: Estimated performance (95% CI) for accuracy, specificity, sensitivity, PPV, NPV and F1-score for 3-year risk of recurrent VTE following cessation of therapy using the model developed by Ensor et al, when N=14250 and N=3600 using a simulation based approach to calculating the confidence interval*

|  | Sample size | |
|---|---|---|
|  | **N=14250** | **N=3600** |
| **Accuracy** | 0.323 (0.315, 0.331) | 0.323 (0.308, 0.339) |
| **Specificity** | 0.257 (0.249, 0.265) | 0.257 (0.242, 0.272) |
| **Sensitivity** | 0.923 (0.909, 0.937) | 0.922 (0.895, 0.950) |
| **PPV** | 0.121 (0.115, 0.127) | 0.121 (0.109, 0.133) |
| **NPV** | 0.968 (0.962, 0.974) | 0.967 (0.956, 0.979) |
| **F1-score** | 0.214 (0.204, 0.224) | 0.214 (0.195, 0.233) |

*Figure 1: Histogram of Beta(1.33,1.75) distribution*

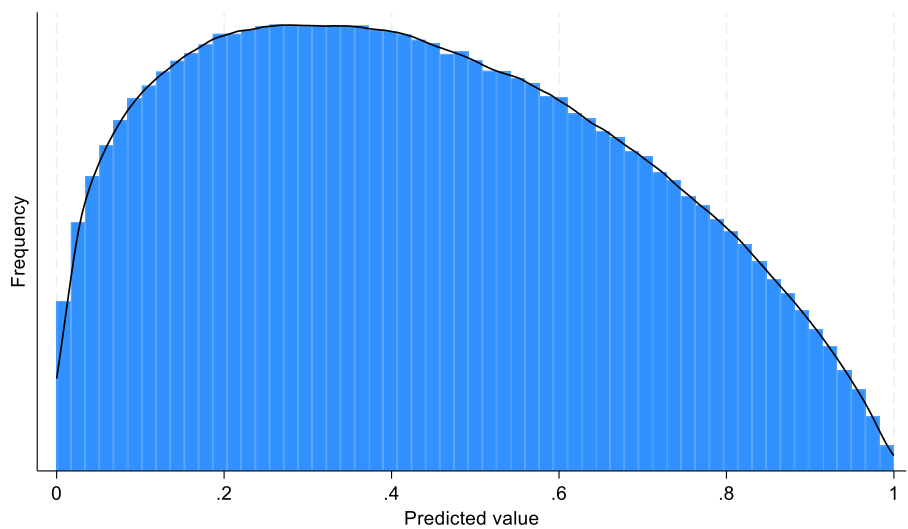